\documentclass[%
 reprint,
 amsmath,amssymb,
 aps,
]{revtex4-2}

\usepackage{times}
\usepackage{graphicx}
\usepackage{dcolumn}
\usepackage{bm}

\begin{document}

\title{Discrete embedded solitary waves and breathers in one-dimensional nonlinear lattices}


\author{Faustino Palmero}
\email[Corresponding author. Electronic address:]{palmero@us.es}
\affiliation{Nonlinear Physics
Group. Escuela T\'{e}cnica Superior de Ingenier\'{\i}a
Inform\'{a}tica. Departamento de F\'{\i}sica Aplicada I. Universidad de
Sevilla. Avda.  Reina Mercedes, s/n. 41012-Sevilla (Spain)}

\author{Mario I. Molina}
\affiliation{Departamento de F\'{\i}sica, Facultad de Ciencias, Universidad de Chile, Las Palmeras 3425, Nunoa, Santiago, Chile}

\author{Jes\'us Cuevas-Maraver}
\affiliation{Grupo de F\'{\i}sica No Lineal, Departamento de F\'{\i}sica Aplicada I,
Universidad de Sevilla. Escuela Polit\'{e}cnica Superior, C/ Virgen de Africa, 7, 41011-Sevilla, Spain}
\affiliation{Instituto de Matem\'{a}ticas de la Universidad de Sevilla (IMUS). Edificio
Celestino Mutis. Avda. Reina Mercedes s/n, 41012-Sevilla, Spain, Avda Reina Mercedes s/n, E-41012 Sevilla, Spain}

\author{Panayotis G.\ Kevrekidis}
\affiliation{Department of Mathematics and Statistics, University
of Massachusetts, Amherst, Massachusetts 01003-4515, USA}

\date{\today}

\begin{abstract}
For a one-dimensional linear lattice, earlier work has shown how to systematically construct a slowly- decaying linear potential bearing a localized eigenmode embedded in the continuous spectrum. Here, we extend  this idea in two directions: The first one is in the realm of the discrete nonlinear Schrodinger equation, where the linear operator of the Schrodinger type is considered in the presence of a Kerr focusing or defocusing nonlinearity and the embedded linear mode is continued into the nonlinear regime as a discrete solitary wave. The second case is the Klein-Gordon setting, where the presence of a cubic nonlinearity leads to the emergence of embedded-in-the-continuum discrete breathers. In both settings, it is seen that the stability of the modes near the linear limit turns into instability as nonlinearity is increased past a critical value, leading to a dynamical  delocalization of  the solitary wave (or breathing) state. Finally, we suggest  a concrete experiment to observe these embedded modes using a bi-inductive electrical lattice.

\end{abstract}


\maketitle


\section{\label{sec:level1}  Introduction}
In quantum mechanics it is typically expected that for a particle in the presence of an external potential (whose value at infinity is set as the zero level), the eigenstates consist of either  extended modes, via a quasi-continuum spectrum with positive energies, or of localized modes, with negative energy, forming a discrete (so-called point) spectrum. This result, which  stems from a general analysis of the Schr\"odinger equation, was challenged by Wigner and von Neumann~\cite{Wigner} who showed that it was possible for a localized state to co-exist with the extended modes inside the continuum band. In a seminal work, they constructed explicitly a three-dimensional potential, using a reverse-engineering approach, to produce a setting that was tailored to support a so-called bound state in the continuum or embedded mode (EM). The method had to face some difficulties since, for instance, the EM produced decreased in space as a power law, making it non-renormalizable. Ideally, one would like to have a {\em bona fide} normalized, localized mode that is decoupled from the continuum, like a resonance with zero width.

Regarded at first as a mathematical curiosity, the topic of EM has re-emerged recently with an abundance of works on its theory and applications in many diverse areas where wave phenomena (quantum and classical) are dominant. During the 1970s, Stillinger~\cite{stillinger1,stillinger2} and Herrick~\cite{herrick} improved the theory of Wigner and von Neumann and  pointed out that EMs could be found in certain atomic and molecular systems. They also suggested the use of superlattices to construct potentials that could support EMs~\cite{superlattice1,superlattice2}. More recently, direct observations of electronic bound states above a potential well and localized by Bragg reflections, were carried out using semiconductor heterostructures~\cite{capasso}. A different approach comes from the physics of resonant states in quantum mechanics. These states are spatially localized  but with non-decaying tails and energies inside the band, and they eventually decay, i.e., they possess a finite lifetime. However, it is possible to arrange conditions in order to make a given resonance interfere negatively with another one and produce a zero width resonance, that is, an EM. This has been shown to occur in a hydrogen atom immersed in a magnetic field, modeled as a system of coupled Coulombic channels, where interference between resonances belonging to different channels can lead to the creation of EM~\cite{coulomb1,coulomb2}. In recent times, EMs have been observed in mesoscopic electron transport and quantum waveguides~\cite{electron1,electron2,electron3,electron4,electron5,electron6,electron7,electron8,electron9}, and in quantum dot systems~\cite{orellana1,orellana2,orellana3,orellana4,orellana5}. A common theme in all the above systems is that the onset of an EM can be traced back to the destruction of the discrete-continuum decay channels by quantum interference effects. The ultimate origin of the EM phenomenon  is regarded nowadays as the result of interference and thus it should be inherent to any wavelike theory besides quantum mechanics. 
The original approach of Wigner and von Neumann has been extended to the case of a discrete and periodic system, such as the usual tight-binding model. There, the methodology consists of choosing an envelope that modulates a given chosen extended eigenstate, and then imposing that its energy coincides with the original one (no envelope). To achieve this a site energy distribution is chosen judiciously, so that the energy for the modulated state  coincides with the energy of the original mode~\cite{1D1,1D2,2D}; see also
details below.

While the above ideas have been developed in the linear realm, a parallel
track of efforts has concerned a wide range of nonlinear models bearing
localized modes, as summarized in a number of reviews~\cite{reviews}. 
The corresponding applications of the above studies are broad and far-reaching and
extend from the nonlinear optics of waveguide arrays~\cite{moti}
to the dynamics of atomic Bose-Einstein condensates in periodic
potentials~\cite{bec_reviews} and from the micro-mechanical models of
cantilever arrays~\cite{sievers} and electrical lattices~\cite{remoissenet}
to prototypical models of
the DNA double strand~\cite{peyrard}. Two of the most canonical
mathematical sets of models where localized nonlinear waveforms
emerge consist of the discrete nonlinear Schr{\"o}dinger (DNLS)
model~\cite{dnls} and the nonlinear discrete variant of the wave
equation in the form of
nonlinear Klein-Gordon lattices~\cite{reviews,dp4}. These represent
generic frameworks featuring the interplay of discreteness,
dispersion and nonlinearity and are in one or another variant
(i.e., for different nonlinearities, external potentials etc.)
involved in the description of each one of the above physical settings.

In this work we combine these two aspects of dynamical lattices.
More specifically, we
study the problem of constructing an embedded mode (EM) inside the band of three different, but related systems: (a) A nonlinear DNLS lattice
(section II) (b) A $\phi^4$ chain (section III) (c) A one-dimensional bi-inductive electrical  lattice (section IV).
Our aim is, upon construction of the linear EM mode, to continue the relevant
mode in the nonlinear realm, either as a solitary wave (in the DNLS setting)
or as a discrete breather (in the $\phi^4$ and electrical lattice cases)
and continue it parametrically over frequencies. We observe that the relevant
modes can be continued within the (discretized ---due to the finiteness of
the domain---) continuous spectrum, yet their stability changes along
this continuation. We detail the relevant instabilities and the
corresponding bifurcation diagrams. We also dynamically
monitor the instability evolution, as well as propose 
an experimentally realistic setting for the implementation of
the ideas proposed herein. We believe that the relevant
findings offer a sense of the potential applicability of EMs in
nonlinear dynamical lattices that are of wide relevance in the above
discussed physical settings. Our presentation closes with section V
which summarizes our findings and presents our conclusions, as well
as some interesting directions for further study.
We also note in passing that the EMs obtained here are 
distinct 
from earlier works involving the discretization of a continuous model that, in turn, bears embedded solitons and where the
EMs are genuinely nonlinear entities; an example
of the latter in a (spatially homogeneous, i.e., involving
no external potential) lattice model with competing quadratic and
cubic nonlinearities has been studied in ref.\cite{malomed}. 

\section{\label{sec:DNLS}The DNLS chain}

We start by considering a discrete model described by the DNLS equation with a
set of impurities following the ideas of~\cite{1D1,1D2,2D} in the form:

\begin{equation}
i \dot{\Psi}_n + \gamma |\Psi_n|^2 \Psi_n + C(\Psi_{n+1}+\Psi_{n-1})+\epsilon_n \Psi_n=0,
\label{DNLS}
\end{equation}
where $n=1 \dots N$, while $N$ represents
the size of the system. We consider fixed-end boundary conditions, and $\epsilon_n$
represents the effective external potential leading to the existence
of the EM. As is customarily the case in such DNLS lattices~\cite{dnls},
the energy
$H=\sum (C/2) |\Psi_{n+1}-\Psi_n|^2 -(\gamma/2) |\Psi_n|^4 + \epsilon_n
|\Psi|^2$ and the power $P=\sum |\Psi_n|^2$ are conserved magnitudes. Stationary states in such DNLS settings arise in the form of
standing waves: $\Psi_n=\phi_n \exp(i \omega t)$, with $\phi_n$ being
the solutions of the steady state problem:

\begin{equation}\label{eq:DNLS_stat}
(-\omega +\epsilon_n) \phi_n + \gamma \phi_n^3+C(\phi_{n+1}+\phi_{n-1})  =  0
\end{equation}
and $\omega$ are the corresponding frequencies.

\subsection{Linear case}

In order to get the effective potential \{$\epsilon_n$\} necessary for the existence of the EM, we follow the prescription of~\cite{1D1,1D2,2D}, consisting in selecting an arbitrary eigenmode \{$\phi'_n$\} with eigenfrequency $\omega'$ of the linear homogeneous lattice (i.e. $\gamma=0$, $\epsilon_n=0\ \forall n$ in (\ref{eq:DNLS_stat})). Then, it is possible to build a spatially localized linear state around site $n_0$ in the linear non-homogeneous case with the same frequency $\omega'$, by choosing an effective potential profile \{$\epsilon_n$\} as:
%
\begin{equation}
    \epsilon_n  = \omega'-C \left[ \left(\frac{f_{n+1}}{f_n}\right) \left(\frac{\phi_{n+1}'}{\phi_n'}\right)+\left(\frac{f_{n-1}}{f_n}\right)
    \left(\frac{\phi_{n-1}'}{\phi_n'}\right)\right],
\end{equation}
where $\{f_n\}$  is a decreasing function with a maximum at $n_0$ and defined to avoid singularities in $\{\epsilon_n\}$  as
\begin{eqnarray}
f_n & = & \prod_{m=1}^{|n-n_0|-1} (1-\delta_{m}) \qquad  n   \neq  n_0, \nonumber \\
f_n & = & 1  \qquad n   =  n_0,
\end{eqnarray}
and
\begin{equation}
\delta_n=\frac{a}{1+|n-n_0|^b} N^2\phi^{'2}_n\phi^{'2}_{n+1}.
\end{equation}
for suitable parameter $a$ and decay exponent $b$. In what follows, we have chosen $a=1/2$ and $b=3/4$. Notice that the effective potential and the EM profiles depend on the particular choice of the homogeneous eigenmode \{$\phi'_n$\}.

\begin{figure}[h]
	\centering
	\includegraphics[width=\columnwidth]{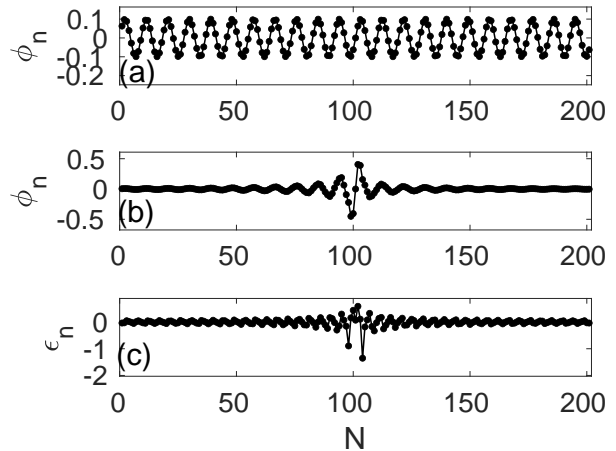}
	\caption{ Linear case: linear extended state in the homogeneous case (a) and linear localized state in the inhomogeneous case (b), with the latter being constructed in line
	with the prescription analyzed in the text. Spatial distribution of impurities $\epsilon_c$ is shown in (c). Here, $N=201$, $C=1$, $n_0=101$, $\omega'=1.5497$.}
	\label{lineal_DNLS}
\end{figure}

\begin{figure}[h]
	\centering
	\includegraphics[width=\columnwidth]{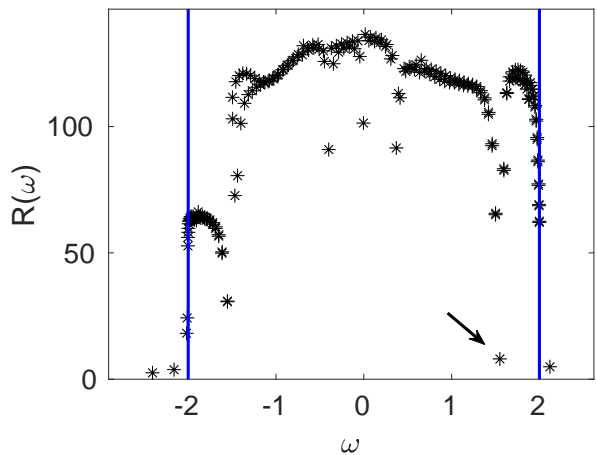}
	\caption{Linear case: Participation ratio $R$ as a function of the frequency corresponding to eigenstates in the inhomogeneous system. Vertical (blue) lines show boundary limits corresponding to the phonon band in the homogeneous chain. The localized state around $n_0$ is shown with an arrow. $N=201$, $C=1$, and $n_0=101$.}
	\label{loca_DNLS}
\end{figure}

In Figure \ref{lineal_DNLS}  we can see the spatial profiles of the linear state in the homogeneous case \{$\phi'_n$\}, of frequency $\omega'=1.5497$, that we used to build the effective potential, the EM with the same frequency $\omega'$ and the spatial distribution of impurities $\epsilon_n$.
As expected, the EM is the only localized mode inside the band  $\omega \in [-2C, 2C]$ of linear modes of the homogeneous lattice. This can be confirmed through Fig. \ref{loca_DNLS}, where the participation ratio $R$ on the vertical axis is defined as
\begin{equation}
R=\frac{\left(\sum_n |\phi_n|^2\right)^2}{\sum_n|\phi_n|^4},
\end{equation}
with $\phi_n$ being the eigenmode of the linear non-homogeneous equation (\ref{eq:DNLS_stat}). A large value of $R$, comparable to the lattice size, illustrates that the relevant mode is extended, while a small value of $R \ll N$ indicates the presence of localization. We note the existence of some states outside the linear modes band of the homogeneous lattice, which possess a participation ratio even smaller than the EM. 
These modes are localized around $n_0$, constituting impurity modes similar to e.g. those studied at \cite{Guillaume}. As the present paper is devoted to EMs (i.e. intraband modes) and out-of-band modes have been 
thoroughly studied, we will not pay attention to them in what follows. In fact, one can to get rid of those impurity modes by devising an effective potential with smaller spatial fluctuations.

\begin{figure}[h]
	\centering
	\includegraphics[width=\columnwidth]{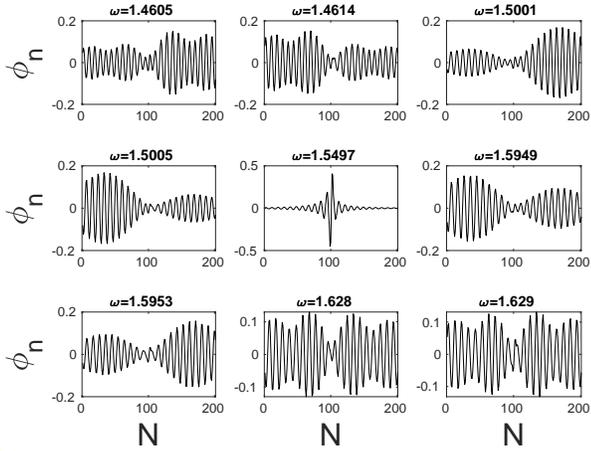}
	\caption{Linear case: States in the band closest in frequency to the localized mode (central panel). $N=201$, $C=1$, $n_0=101$, $\omega'=1.5497$. Notice,
	indeed, the localization at that frequency in comparison with the delocalized
	nature of all neighboring eigenfunctions.}
	\label{linear}
\end{figure}

A numerical examination of the density of states does show the presence of the above mentioned impurity states outside the band, but it does not show any gaps, implying that the mode that has been accordingly prescribed based on the above methodology is indeed inserted inside the continuous band and is the only mode inside the band that is found to be localized. This is confirmed by examining the spatial profiles of the modes.
Indeed, in Fig. \ref{linear} we show the spatial profiles of the EM mode and the closest frequency states in the band.
Nevertheless, because of linearity, the spatial localization occurs for a single (i.e., isolated) frequency value. The idea is to include nonlinear effects and to find a nonlinear localized mode in a range of frequencies around $\omega'$,
i.e., to formulate a continuation problem that allows us to find such nonlinear modes for a wide range of frequencies around the corresponding linear ``bifurcation point''.

\subsection{Nonlinear case}

\begin{figure}[h]
	\centering
	\includegraphics[width=\columnwidth]{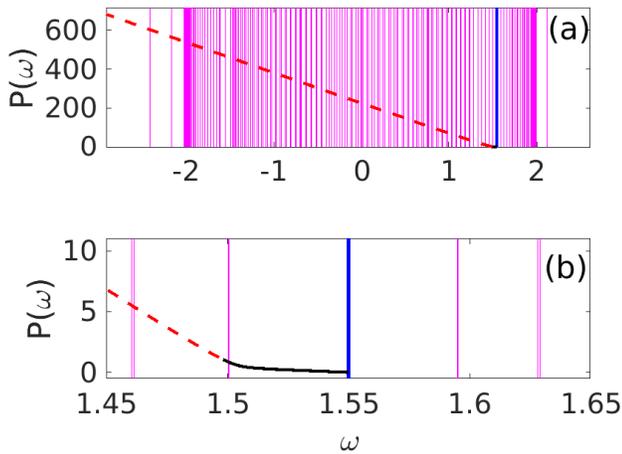}
	\caption{Bifurcation diagram (norm vs frequency) corresponding to nonlinear localized modes in the phonon band  for the defocusing case of $\gamma=-1$. The black line corresponds to the stable solution and the red dashed line to the unstable one emerging from the linear modes. Vertical cyan lines represent linear modes  and the vertical blue line the one corresponding to $\omega'$. (a) Full picture and (b) zoom around $\omega'$, where only linear modes closest to $\omega'$ are shown. $N=201$, $C=1$, $n_0=101$, and $\omega'=1.5497$.}
	\label{nl_band_defocusing}
\end{figure}

\begin{figure}[h]
	\centering
	\includegraphics[width=\columnwidth]{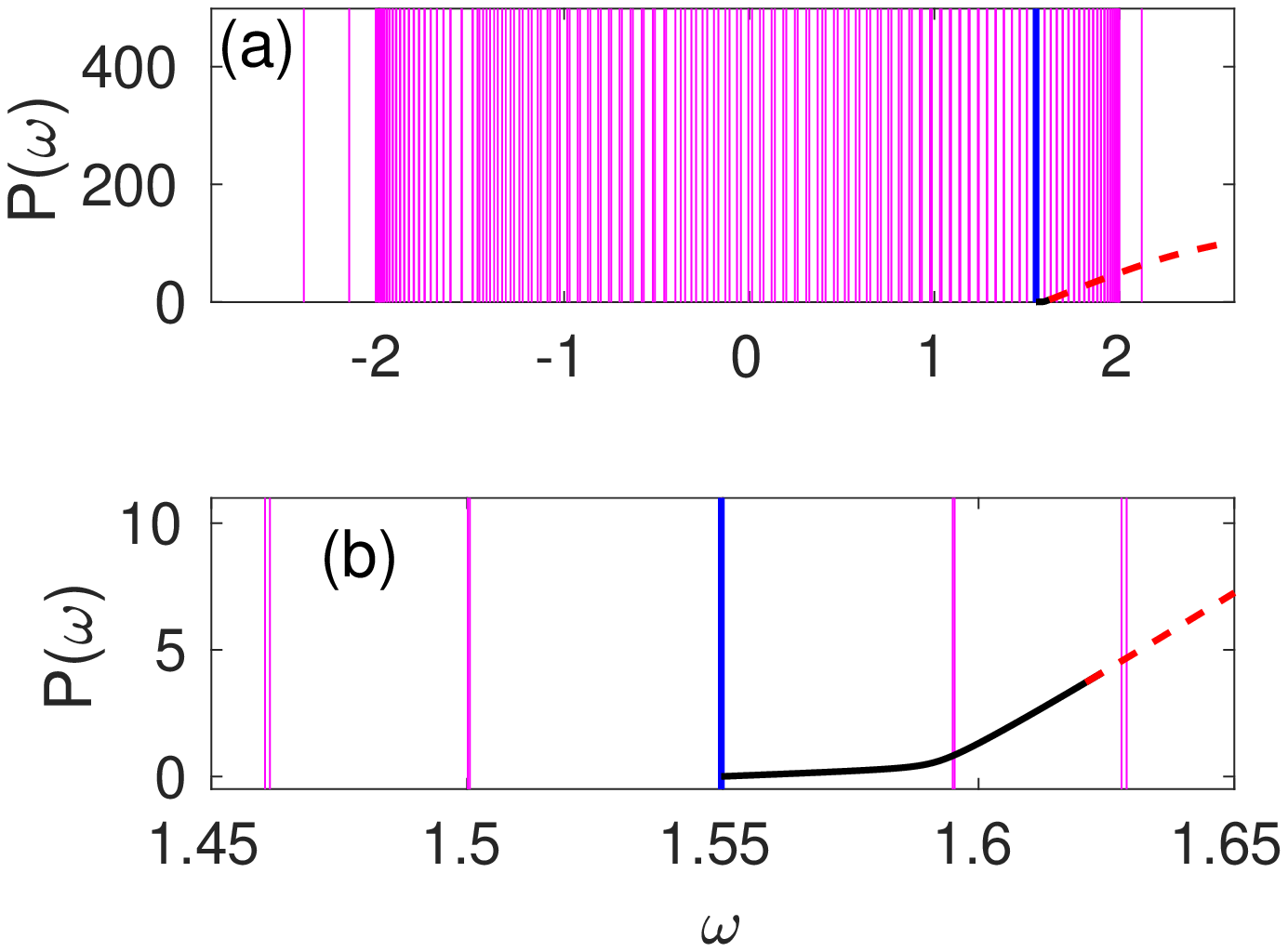}
\caption{Bifurcation diagram (norm vs frequency) corresponding to nonlinear localized modes in the phonon band  for the focusing case of $\gamma=1$.
The black line corresponds to the stable solution and the red dashed line to the unstable one emerging from the linear modes.
Vertical cyan lines represent linear modes  and the vertical blue line the one corresponding to $\omega'$. (a) Full picture and (b) zoom around $\omega'$, where only linear modes closest to $\omega'$ are shown. $N=201$, $C=1$, $n_0=101$, and $\omega'=1.5497$.}
\label{nl_band_focusing}
\end{figure}

\begin{figure}[h]
	\centering
	\includegraphics[width=0.45\columnwidth]{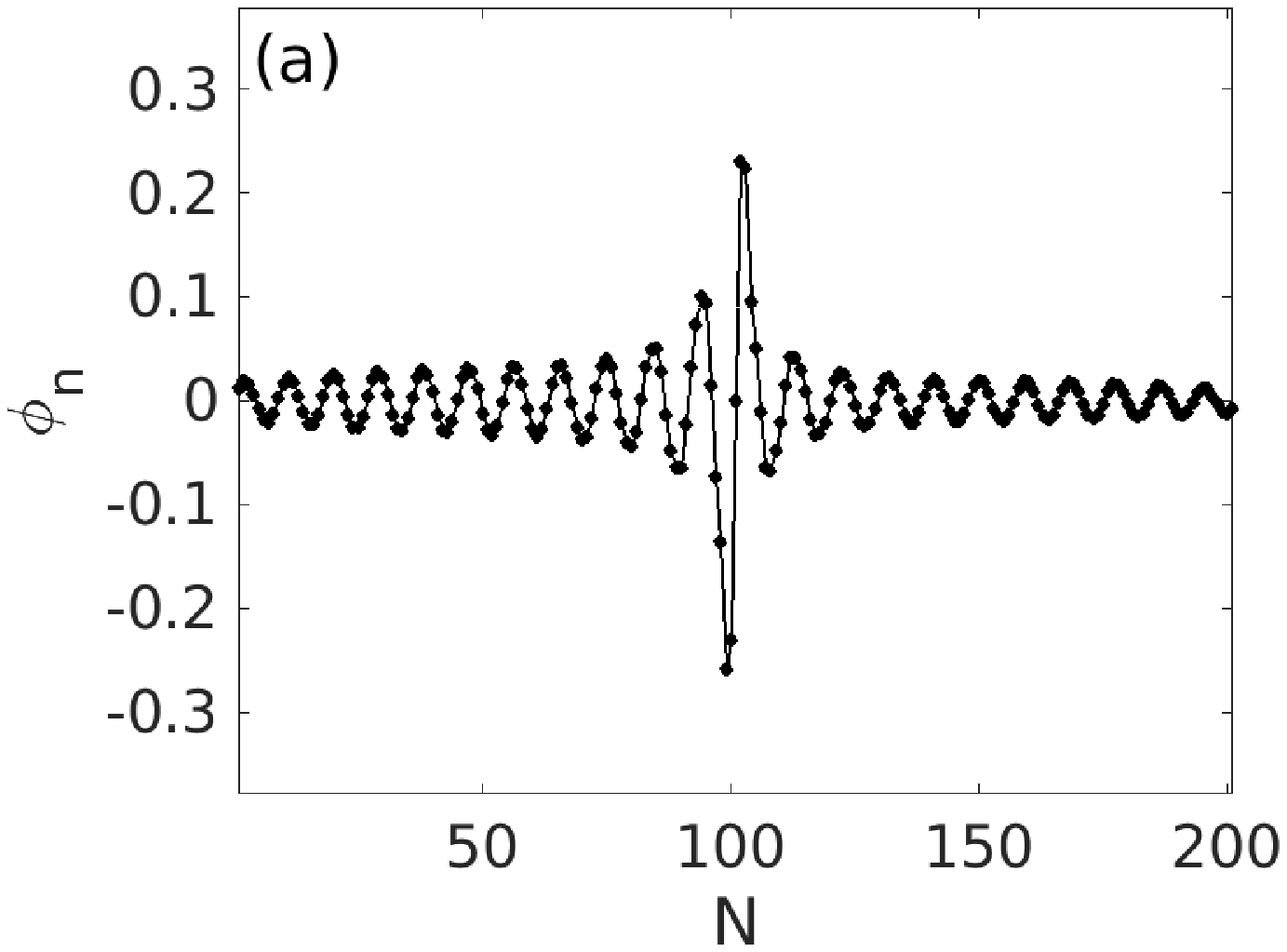}
	\includegraphics[width=0.45\columnwidth]{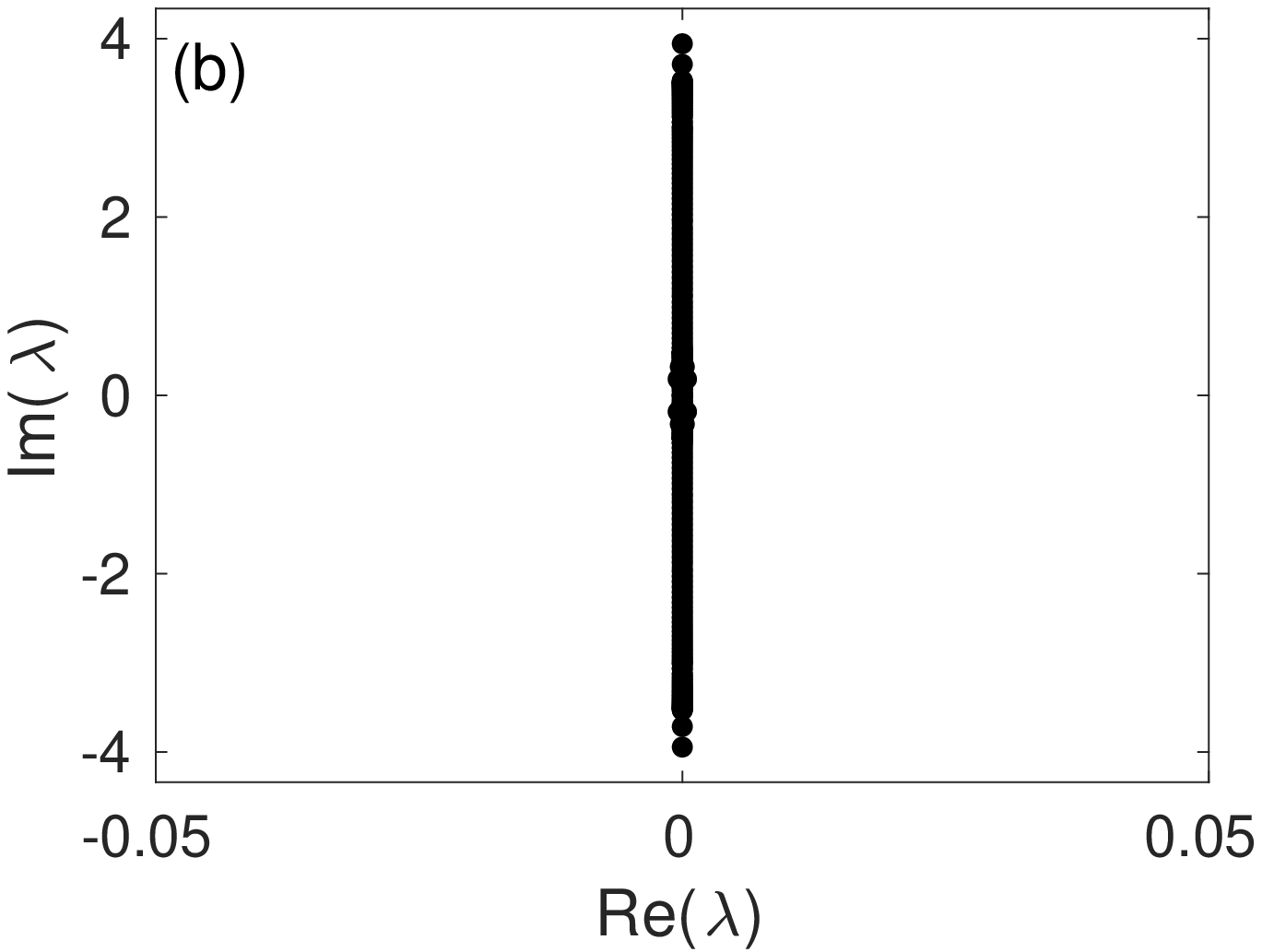}
	\includegraphics[width=0.45\columnwidth]{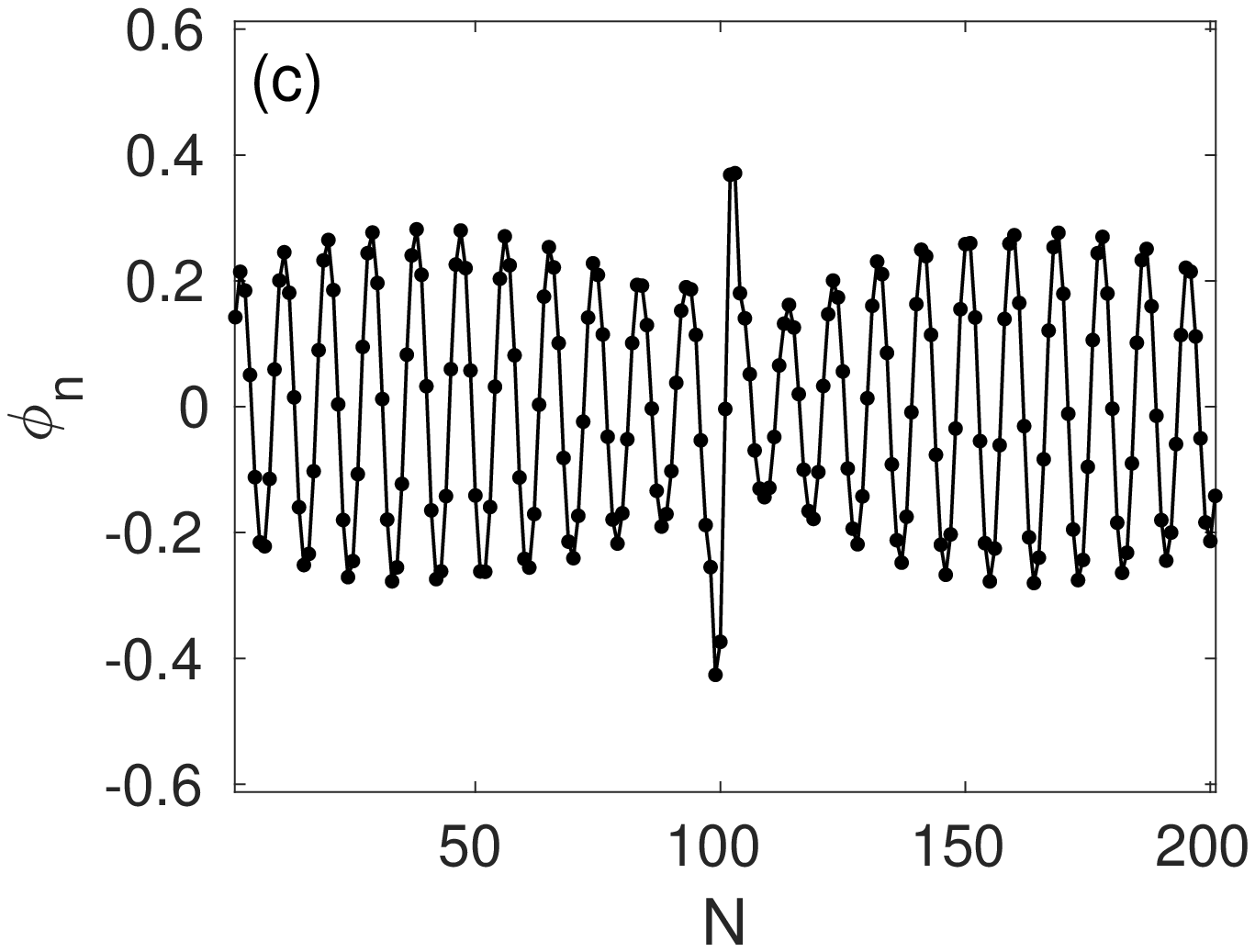}
	\includegraphics[width=0.45\columnwidth]{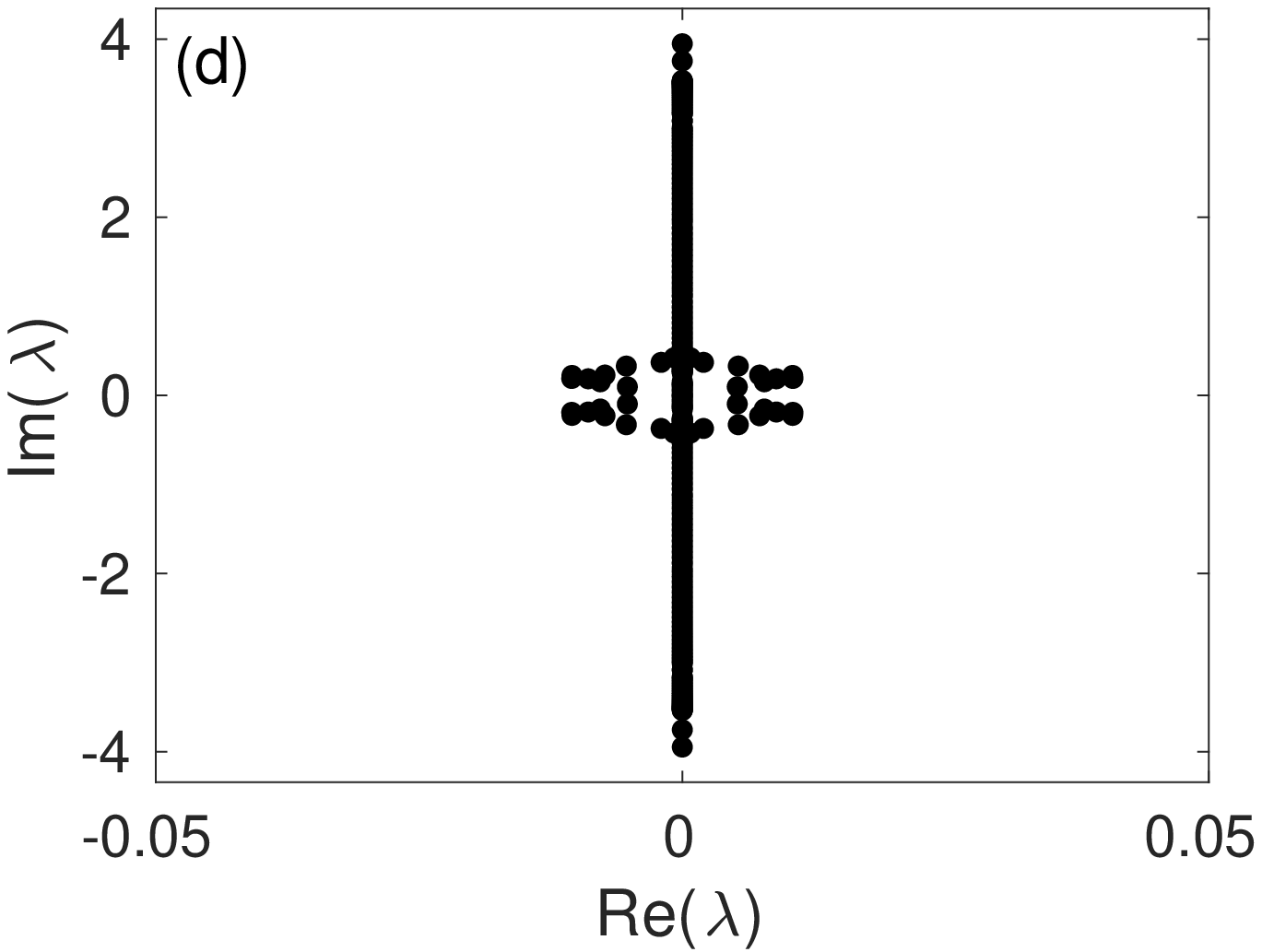}
	\includegraphics[width=0.45\columnwidth]{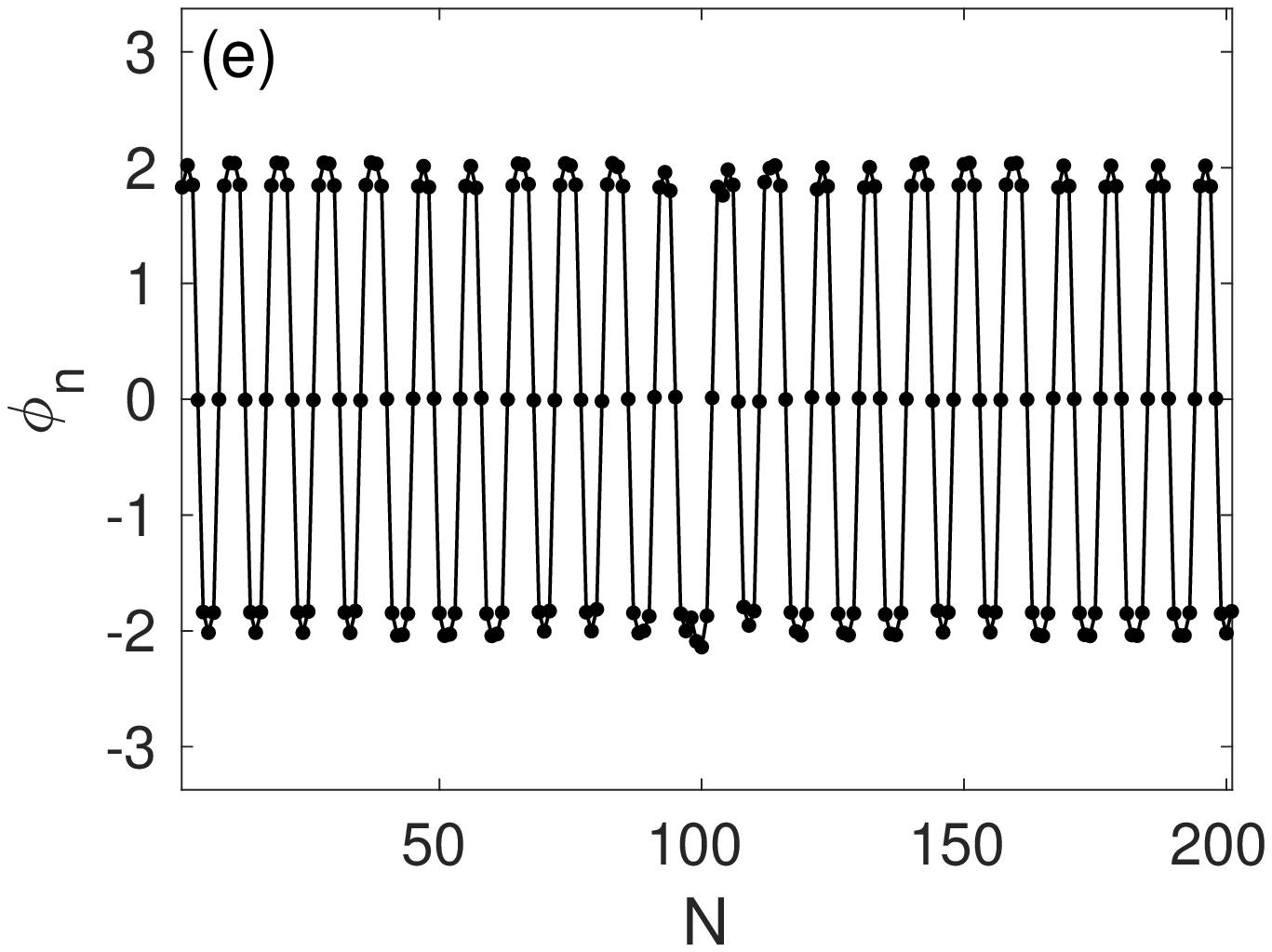}
	\includegraphics[width=0.45\columnwidth]{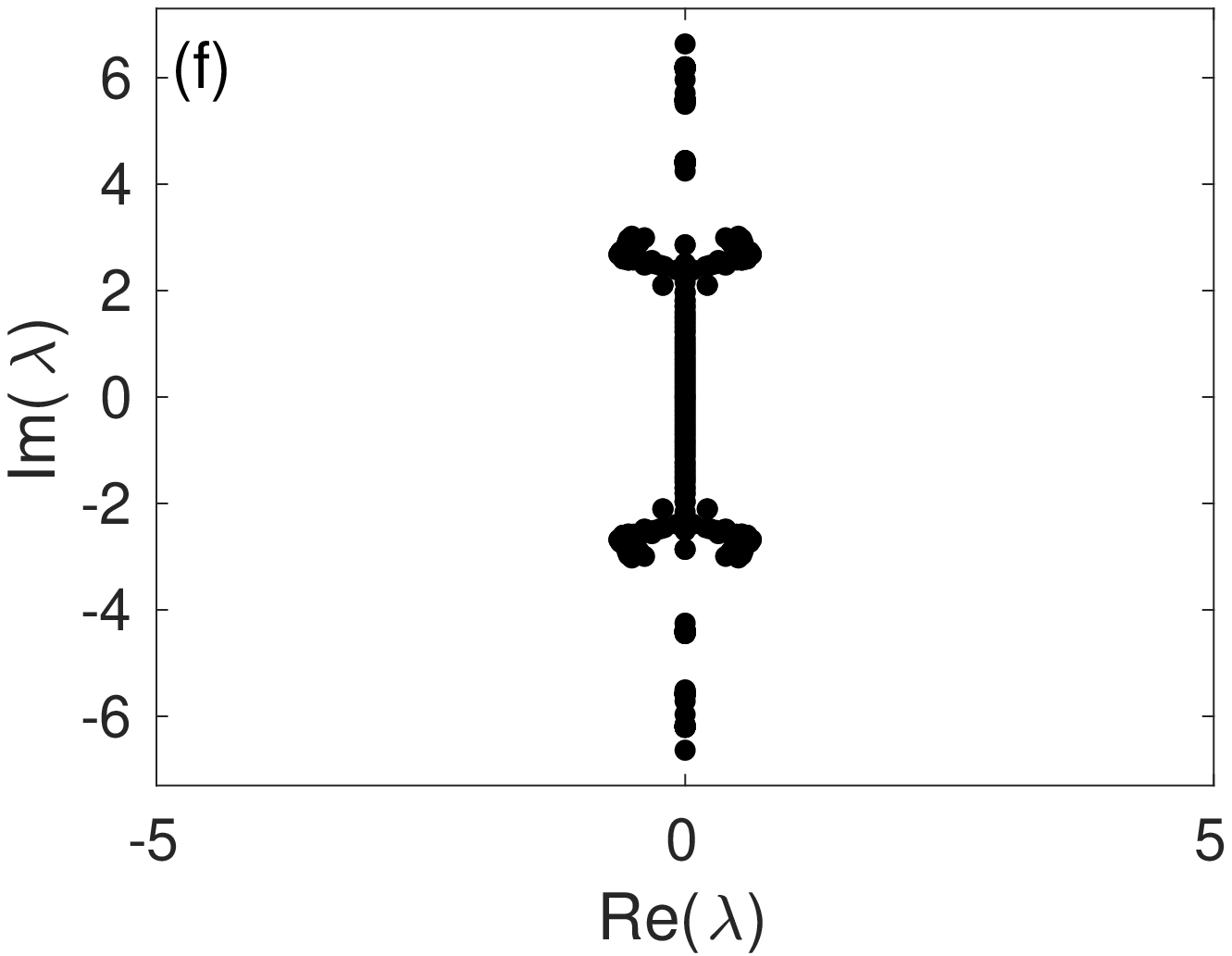}
	\caption{Spatial profiles and linear stability results corresponding to  nonlinear localized excitations in the defocusing case. (a);  stable excitation ($\omega=1.51$); (c) unstable localized excitation ($\omega=1.45$); (e) unstable delocalized state below the phonon band ($\omega=-2.29$). (b), (d) and (f) show their corresponding real and imaginary part of the eigenvalues $\lambda$. Here, $\gamma=-1$, $N=201$, $C=1$, $n_0=101$, $\omega'=1.5497$. Notice the instability of the
	latter two states due to the presence of eigenvalues with 
	non-vanishing real part.}
	\label{spatial_profile_DNLS}
\end{figure}

It has indeed been possible to find the nonlinear spatially localized mode with a spatial profile similar to the linear one and determine its stability by means of numerical techniques similar to those discussed, e.g.,
in~\cite{reviews,dnls}. In general, we have found that small amplitude stable nonlinear localized modes exist for frequencies close to the linear one. When the frequency progressively changes (i.e., detunes further from the linear limit), the amplitude increases and these modes become unstable with a correspondingly more delocalized spatial profile for larger detuning, as shown in Figs. \ref{nl_band_defocusing} and \ref{nl_band_focusing}. On the other hand, Fig.~\ref{spatial_profile_DNLS} shows some examples of the spatial profiles and spectral plane corresponding to stable and unstable embedded solitons.

Notice that the soliton past a given value of the frequency (close to eigenfrequency of a lineal eigenmode), hybridizes with the linear modes and becomes a so-called hybrid soliton (see e.g. \cite{WSL}), i.e. the tails asymptote to non-vanishing amplitude oscillations alike to wings.

\begin{figure}[h]
	\centering
	\includegraphics[width=\columnwidth]{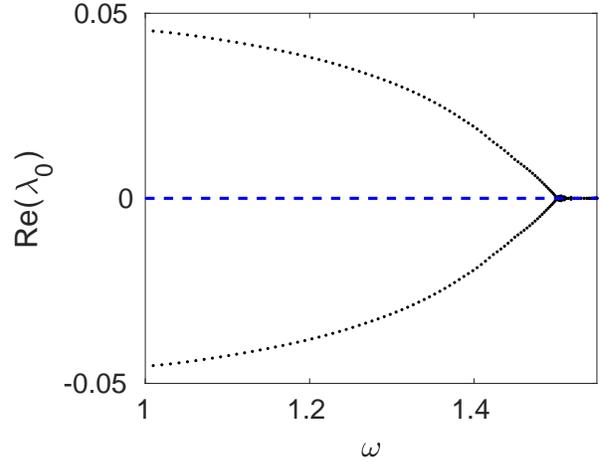}
	\caption{Real part (black dots) corresponding to the
	eigenvalues $\lambda_0$ with the largest real part (in absolute value) of the nonlinear mode emerging from the linear localized mode in the defocusing case ($\gamma=-1$) as a function of the frequency. Due to the symmetries of the system, the eigenvalues appear in quartets (if $\lambda$ is an eigenvalue then so are $-\lambda$ and both of the corresponding complex conjugates). Here,
	$N=201$, $C=1$, $n_0=101$, $\omega'=1.5497$. Notice the critical
	point in the vicinity of $\omega=1.5$, whereby the real part substantially
	increases (as $\omega$ decreases)  mirroring the instability of the relevant
	dynamical branch. The blue dashed line representing $\text{Re}(\lambda_0)=0$ is included for reference.}
	\label{sta}
\end{figure}

The defocusing ($\gamma<0$) and focusing cases ($\gamma>0$) show similar behaviours: in both cases the nonlinear modes emerge from the linear one but in the focusing (defocusing) case, when the frequency increases (decreases), the soliton amplitude increases and the profile becomes more delocalized because of the increment of the wing amplitude. In both cases, the solitons become unstable past a bifurcation point. The relevant destabilization
arising from a Hamiltonian Hopf bifurcation occurs when two pairs of stability eigenvalues $\lambda$ with the same $|\mathrm{Im}(\lambda)|$ collide. Determining the control parameter value corresponding to the bifurcation point is, however, not entirely trivial, as it can be appreciated in Fig. \ref{sta}, because small bifurcations can take place corresponding to instabilities on the wings. We have considered a solution to be stable when $|\mathrm{Re}(\lambda)|<10^{-3}$ for all the eigenvalues of the stability matrix.

\begin{figure}[h]
	\centering
	\includegraphics[width=\columnwidth]{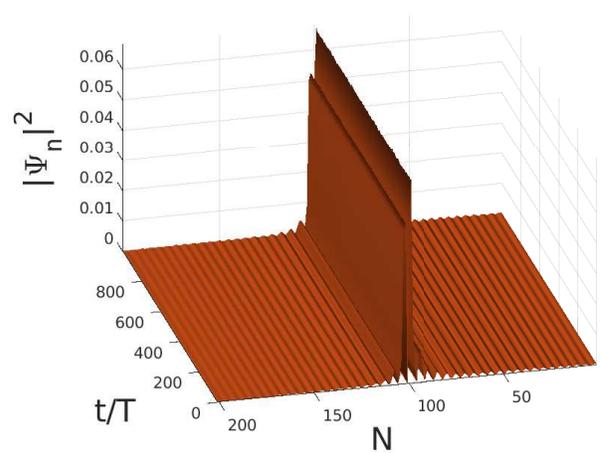}
	\caption{Spatial profiles corresponding to the evolution of a nonlinear localized excitation $\Psi_n(0)=\phi_n+\delta_n$ where $\phi_n$ is the stationary state shown in Fig. \ref{spatial_profile_DNLS} (a), $\omega=1.5101$ and $T=2\pi/\omega$ (stable solutions exist). 
	Shown is the defocusing case  with $\gamma=-1$, $N=201$, $C=1$, $n_0=101$, and $\omega'=1.5497$.}
	\label{profile_DNLS1_evol}
\end{figure}

We have analyzed the effect of a small random perturbation $\delta_n=a_n+ i b_n $, where the real vector $(a_n,b_n)$ components lie in the interval $(-1,1)\times 10^{-3}$. We have numerically integrated equations (\ref{DNLS}) to explore the temporal evolution of a localized excitation
and its dynamical stability properties. More specifically, we consider an initial condition such as $\Psi_n(0)=\phi_n+\delta_n$. As shown in Fig.~\ref{profile_DNLS1_evol},
stable solutions remain localized but unstable ones are delocalized as time evolves. 
On the other hand, unstable ones  eventually manifest their 
respective instability after an initial transient stage.
As a result of the extended nature of the state, we typically observe
the evolution leading to
a rather chaotic waveform with the power $P$ distributed broadly across the lattice sites. Similar results regarding the dynamical redistribution
of the mass of unstable nonlinear states over the entire lattice
have been obtained in the focusing case and hence are not shown here.

\begin{figure}[h]
	\centering
	\includegraphics[width=\columnwidth]{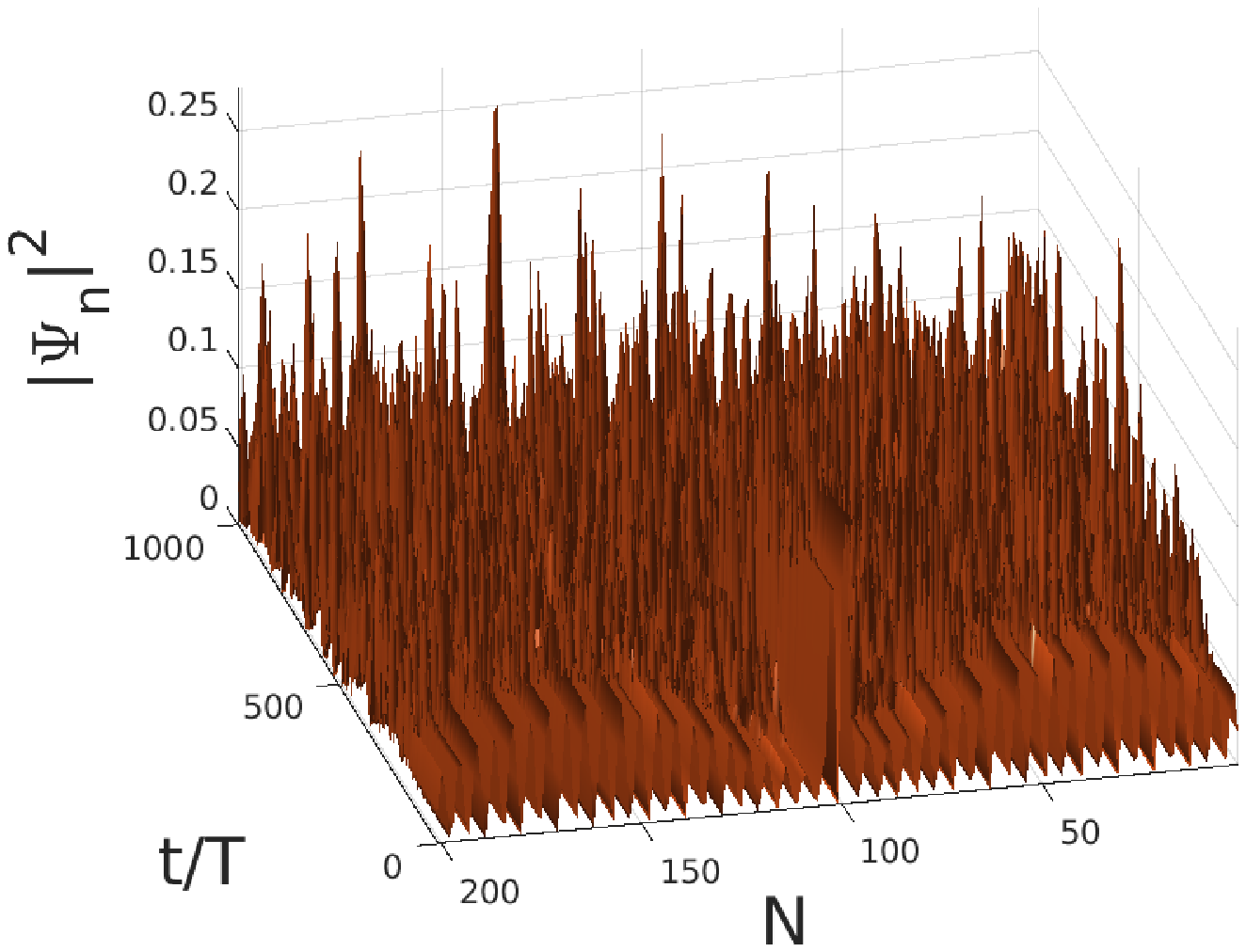}
	\caption{Spatial profiles corresponding to the evolution of a nonlinear localized excitation $\Psi_n(0)=\phi_n+\delta_n$ where $\phi_n$ is the stationary state shown in Fig. \ref{spatial_profile_DNLS} (c), $\omega=1.45$ and $T=2\pi/\omega$ (unstable solutions exist). Defocusing case  $\gamma=-1$, $N=201$, $C=1$, $n_0=101$, and $\omega'=1.5497$. Notice the wide spreading of the relevant waveform, as a result of its instability.}	
	\label{profile_DNLS2_evol}
\end{figure}



	\label{profile_DNLS_fo}



\section{\label{sec:KG}The $\phi^4$ chain}

We now turn to a  $\phi^4$ Klein-Gordon chain described by the equations

\begin{eqnarray}
\ddot{u}_n +  \omega_0^2 u_n+ s u_n^3 + C(2u_n-u_{n+1}-u_{n-1})+\epsilon_n u_n = 0,
\end{eqnarray}
supplemented by free-end boundary conditions. Here  $n  =  1 \dots N$, $N$ is the size of the system. $C$ is the coupling parameter, $\omega_0$ the frequency of the system in the linear limit and $s$ the nonlinear parameter that can be positive (hard potential) or negative (soft potential)~\cite{reviews}. We consider free boundary conditions, and $\epsilon_n$ prescribes the external potential
leading at the linear limit to an EM, and has the same spatial profile
as before.
This system is Hamiltonian with an energy given by
\begin{equation}
H = \sum_n \frac{1}{2} \dot{u}_n^2+\frac{1}{2} (\omega_0^2+\epsilon_n) u_n^2+\frac{1}{4} s u_n^4 + \frac{C}{2}  (u_n-u_{n-1})^2.
\end{equation}
This is the sole conserved quantity of this nonlinear dynamical lattice. Both for this reason, but also because the dependence of $H$ on the
frequency $\omega$ of the breathers has been identified as a key
indicator for their stability~\cite{pelinovsky}, we select to show in
the relevant bifurcation diagrams below  the dependence of $H$ on $\omega$.

\begin{figure}[h]
	\centering
	\includegraphics[width=\columnwidth]{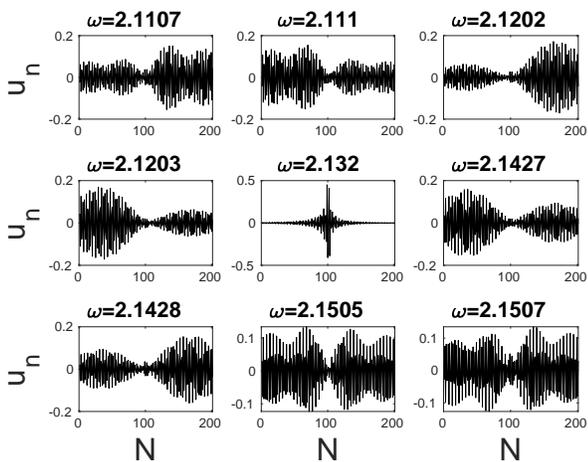}
	\caption{Linear $\phi^4$ chain. Spatial profiles of linear states in the band closest in frequency to the spatially localized mode (central panel). Here, $N=201$, $C=1$, $n_0=101$, $\omega_0=1$, and $\omega'=2.132$.}
	\label{linear_phi4}
\end{figure}

In the homogeneous chain ($\epsilon_n=0$ $\forall n$), it is possible for a set of control parameter values to have stable nonlinear excitations (discrete breathers) with frequencies above the linear frequency band (phonon band) in the hard potential case (of $s=1$) and below this band in the soft case
(of $s=-1$)~\cite{reviews}.
Once again, the ideas of~\cite{1D1,1D2,2D} enable the construction
of an EM at the linear limit, while our aim here is to explore, upon ``sculpting'' an appropriate external potential profile $\epsilon_n$, the possibility of the existence, within the nonlinear problem, of spatially localized states in the
continuous spectrum band with frequencies at or close to the corresponding linear mode. These discrete breathers should, naturally, be reduced to spatially localized linear excitations embedded in the continuous 
spectrum in the linear limit.
An illustration of the latter is shown in Fig.~\ref{linear_phi4},
as constructed by analogy to what was reported previously
for the DNLS (the two linear problems are qualitatively identical).

\subsection{The nonlinear chain}

\begin{figure}[h]
	\centering
	\includegraphics[width=\columnwidth]{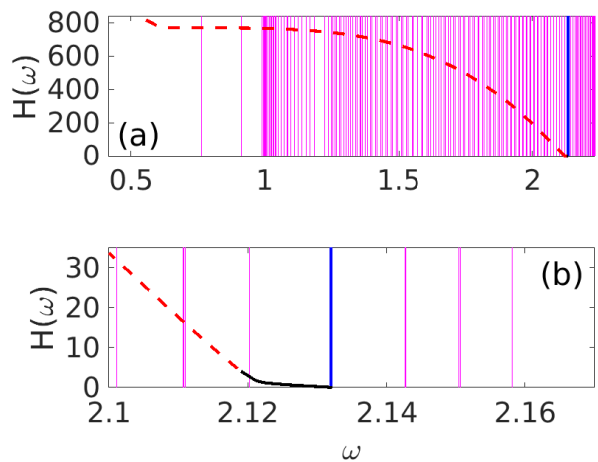}
	\caption{Bifurcation diagram (energy vs frequency) corresponding to nonlinear localized modes in the phonon band  (for the soft potential case
	of $s=-1$). The black line corresponds to the stable solution and the red dashed line to the unstable one emerging from  the linear localized mode. Vertical cyan lines represent linear modes and the vertical blue line the one corresponding to $\omega'$, as shown in Fig. \ref{linear_phi4}. The vertical blue line corresponds to the frequency $\omega'$. (a) Full picture and (b) zoom around $\omega'$, where only linear modes closest to $\omega'$ are shown. $N=201$, $C=1$, $n_0=101$ and $\omega_0=1$.}
	\label{bif_phi4_soft}
\end{figure}

\begin{figure}[h]
	\centering
	\includegraphics[width=\columnwidth]{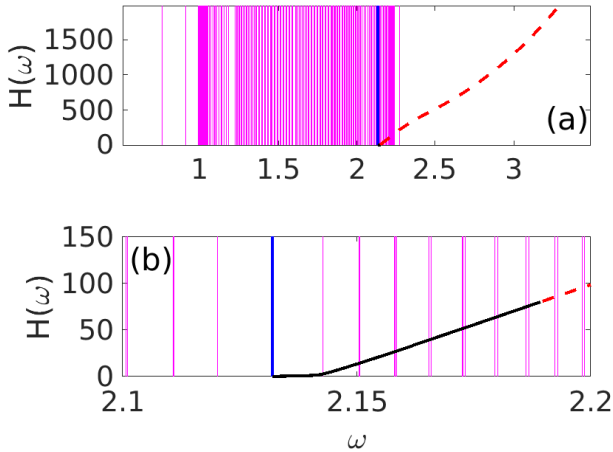}
	\caption{Same as Fig.~\ref{bif_phi4_soft} but for a hard ($s=1$) potential.}
	\label{bif_phi4_hard}
\end{figure}

Using the spatially localized linear state as an initial seed, it is possible to perform numerical continuation to analyze the nonlinear case.
Indeed, we have used the Poincar{\'e} section methodologies analyzed, e.g.,
in~\cite{reviews}, to identify discrete breathers with frequencies progressively
detuning away from the corresponding linear limit.
In general, we have found that the scenario for the resulting
nonlinear time-periodic waveforms is similar to that of
the DNLS chain case. Stable and small amplitude nonlinear localized breathing states emerge from the linear mode and become unstable when the frequency varies and the amplitude increases, as shown in Figs. \ref{bif_phi4_soft} and \ref{bif_phi4_hard}, respectively for the soft potential of $s=-1$ and
the hard potential case of $s=1$. In both scenarios, the immediate
vicinity of the linear mode (with the frequency still within the band)
retains its stability, however sufficient detuning from the linear limit
results in instability, typically via Hamiltonian Hopf bifurcations,
as discussed further below.

\begin{figure}[h]
	\centering
	\includegraphics[width=0.45\columnwidth]{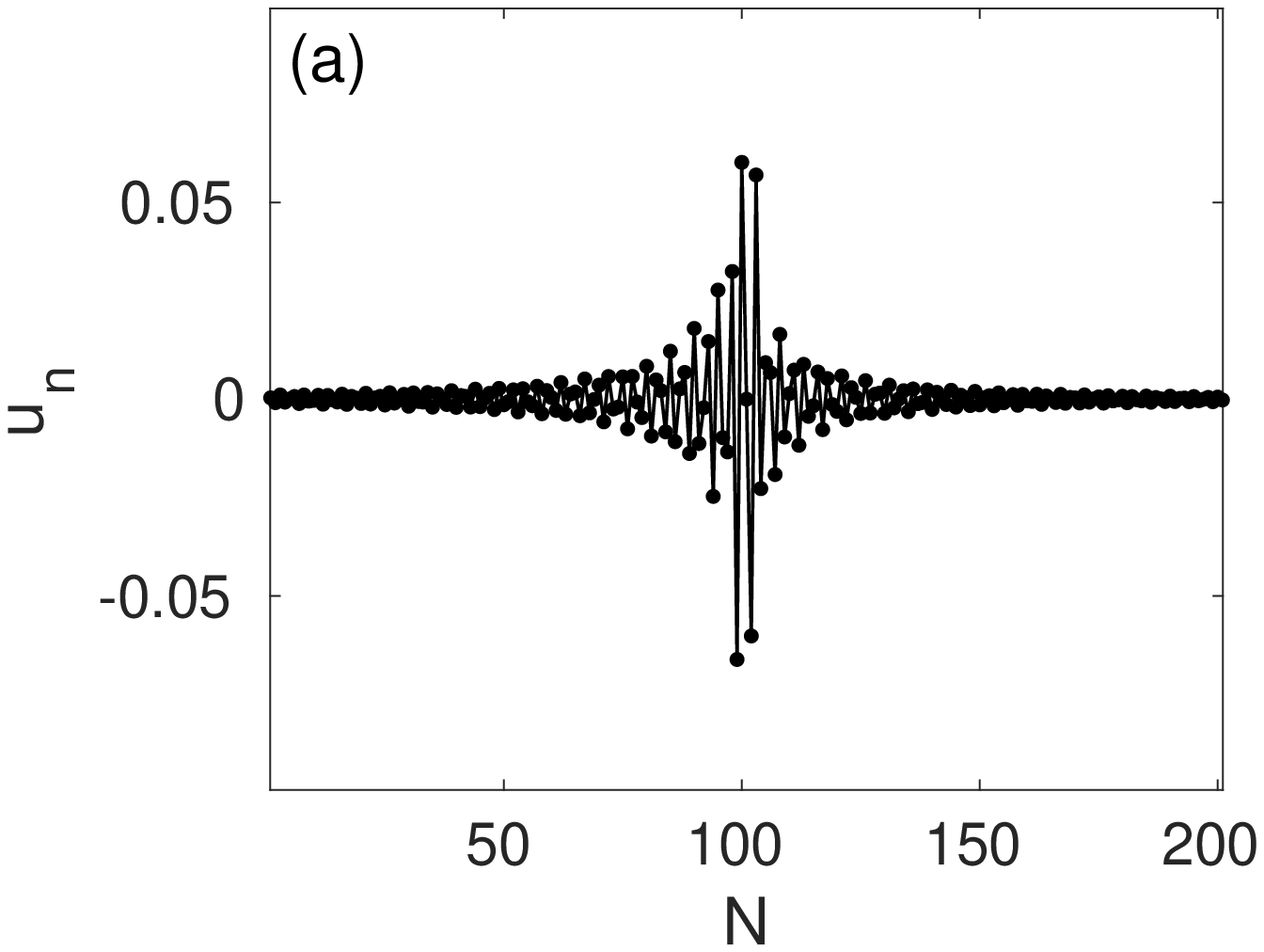}
	\includegraphics[width=0.45\columnwidth]{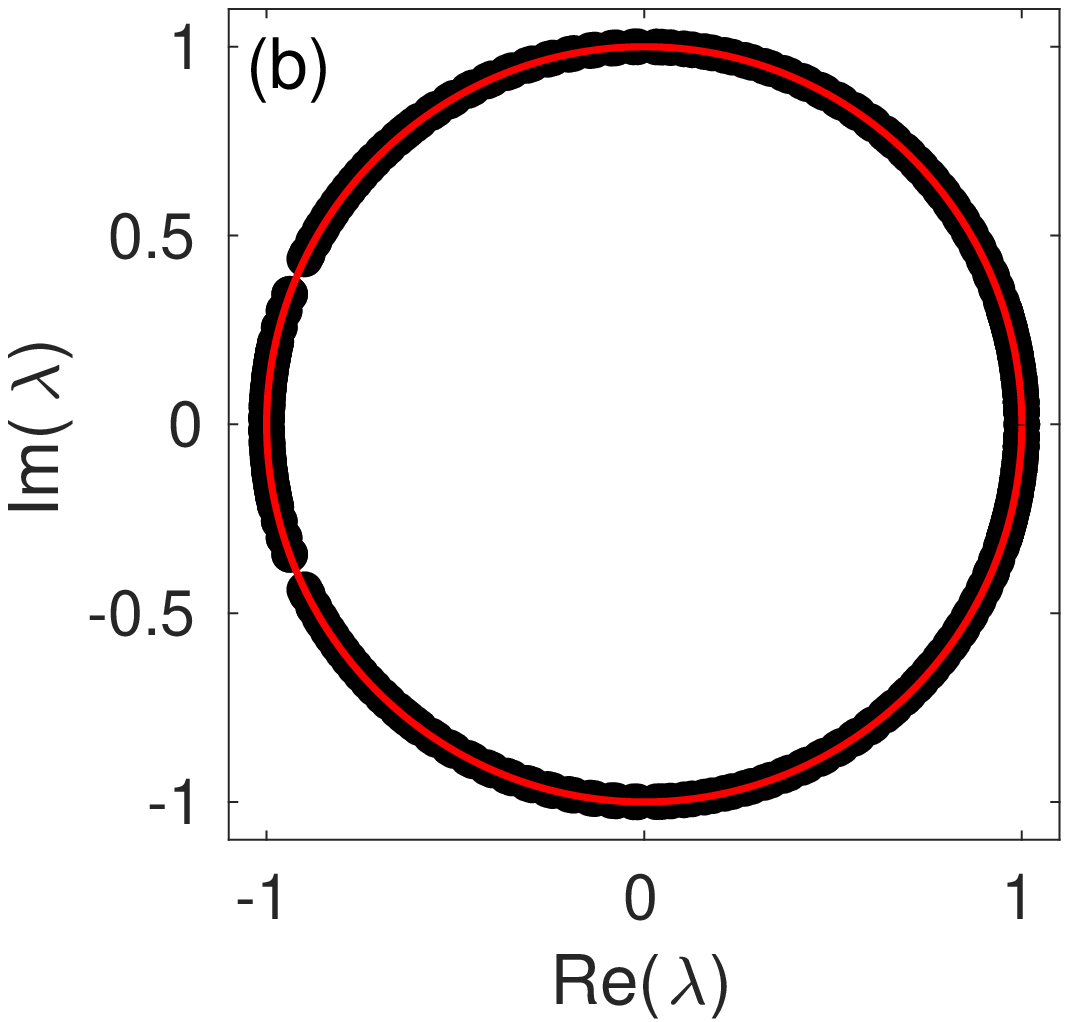}
	\includegraphics[width=0.45\columnwidth]{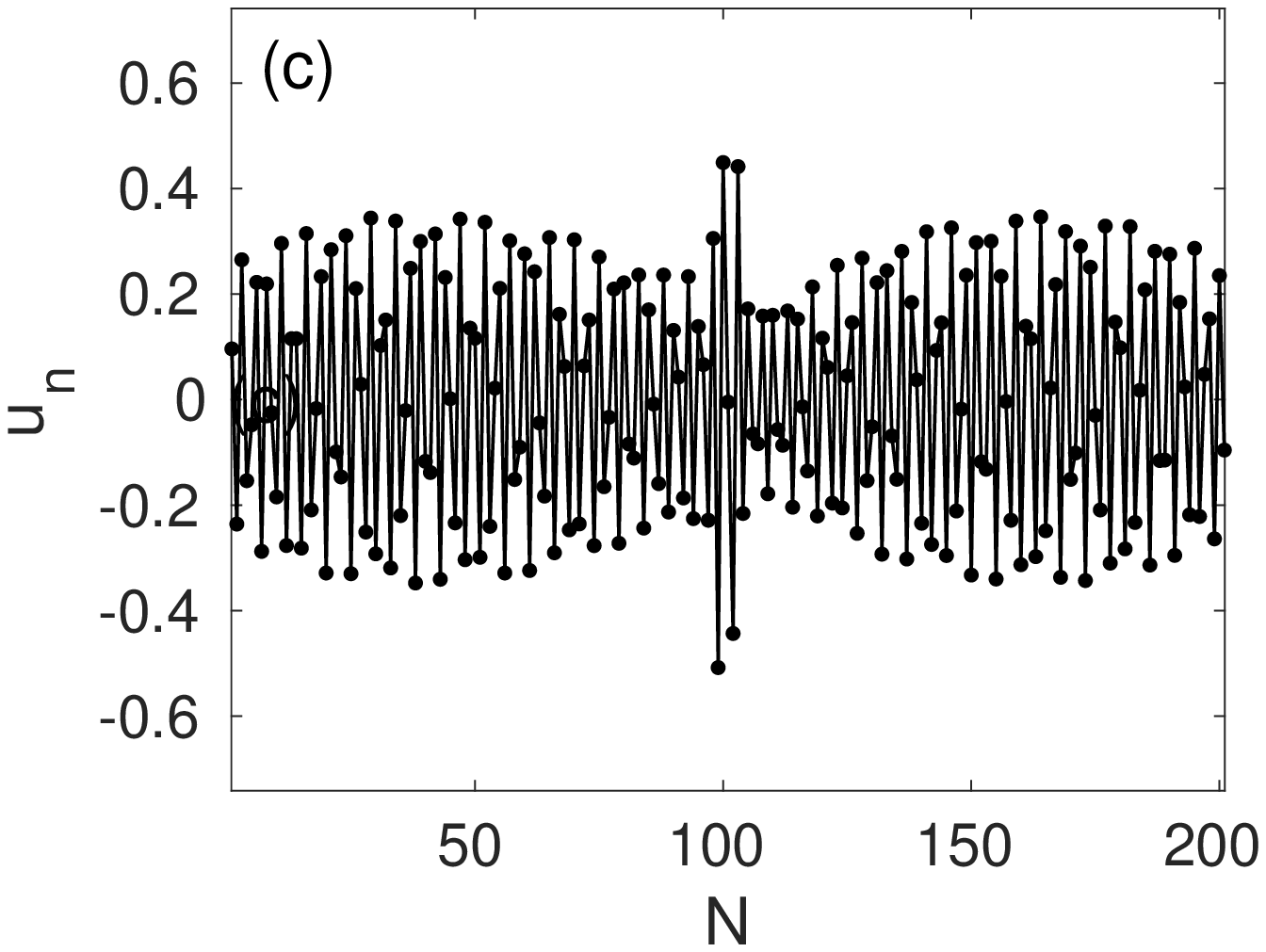}
	\includegraphics[width=0.45\columnwidth]{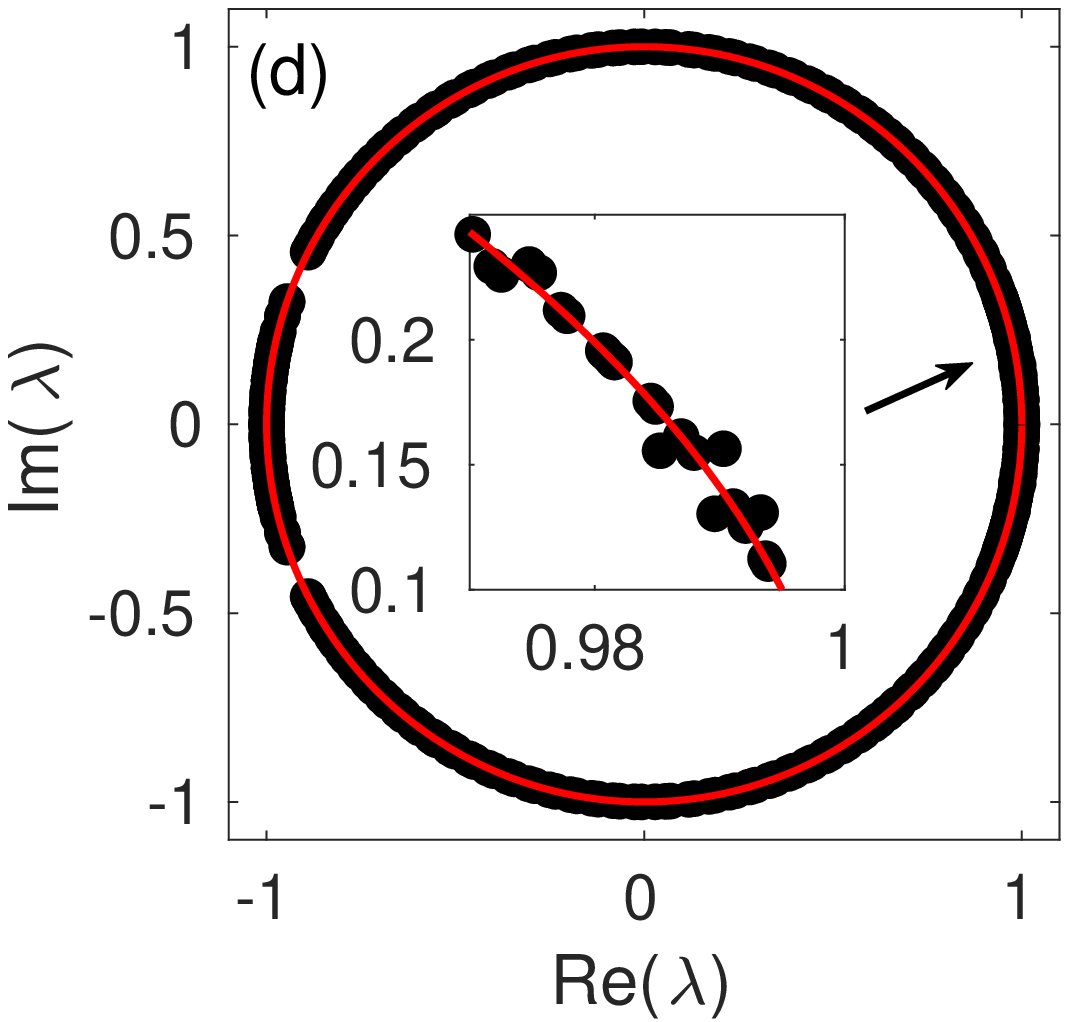}
	\includegraphics[width=0.45\columnwidth]{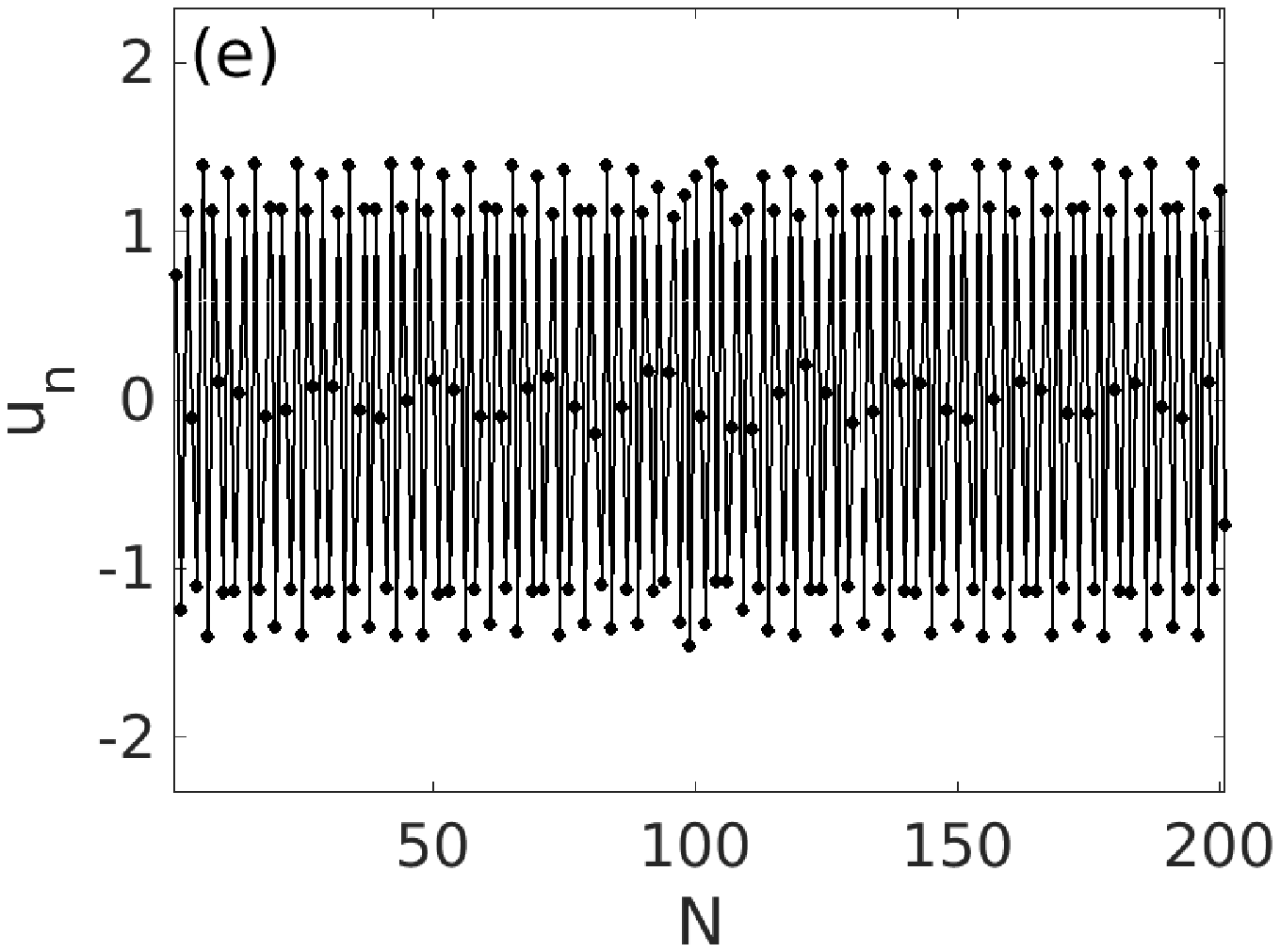}
	\includegraphics[width=0.45\columnwidth]{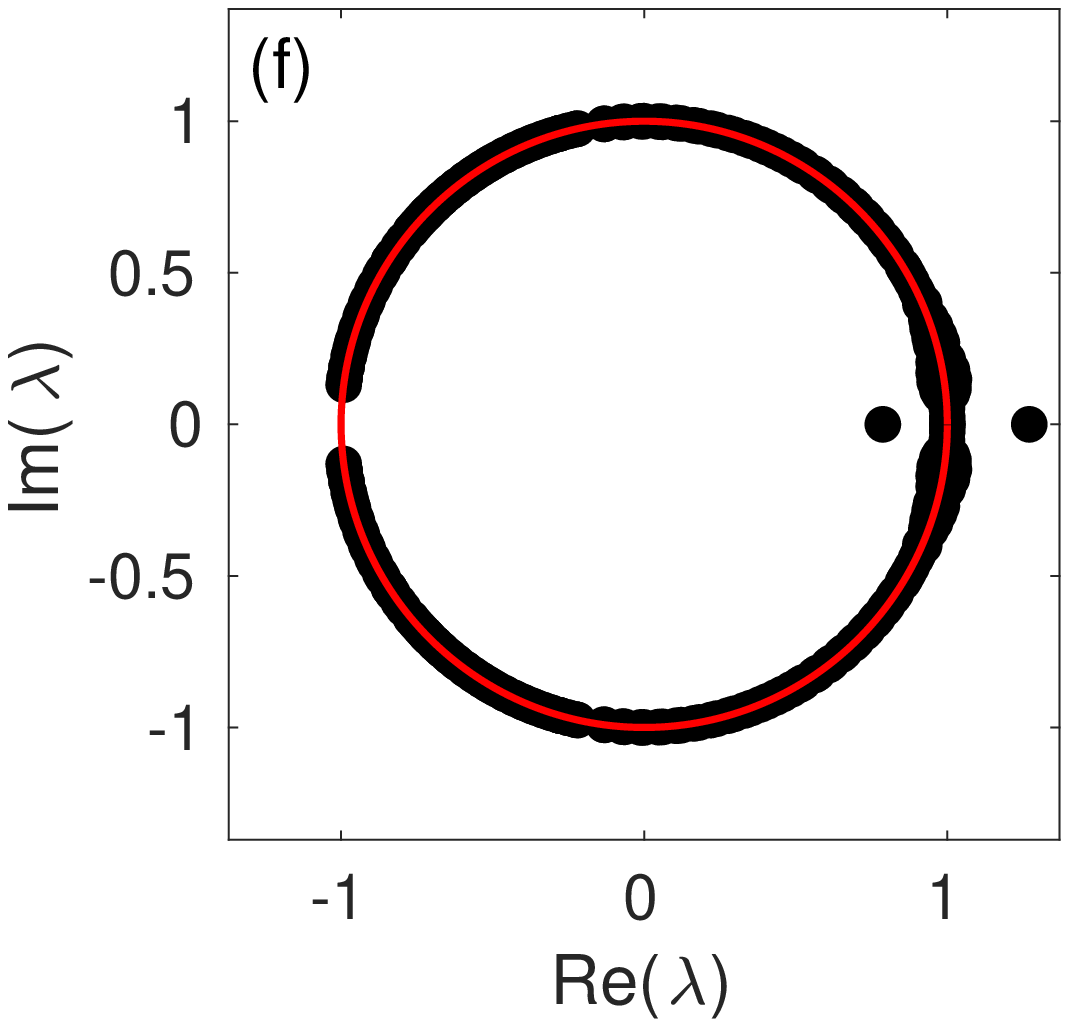}
	\caption{ (a), (c) and (d) Spatial profiles corresponding to  nonlinear localized excitations in the soft potential case. (a);  stable excitation ($\omega=2.131$); (c) unstable localized excitation ($\omega=2.106$); (e) unstable delocalized state  ($\omega=1.805$). (b), (d) and (e) show the corresponding real and imaginary parts of the eigenvalues $\lambda$. Here, $s=-1$, $N=201$, $C=1$, $n_0=101$, $\omega'=2.132$.}
	\label{spatial_profile_KG_soft}
\end{figure}

 In Fig. \ref{spatial_profile_KG_soft} we show the spatial profiles and Floquet spectra corresponding to stable and unstable nonlinear modes in the soft potential case. The presence of a Floquet exponent $\lambda$ with modulus greater than 1 implies instability. When the frequency decreases, 
 oscillatory instabilities involving quartets of Floquet multipliers take place corresponding to (Hamiltonian) Hopf bifurcations, as can be appreciated in Fig. \ref{spatial_profile_KG_soft} (c) and (d). Eventually, lower frequency values produce an exponential (tangent) bifurcation, as shown in Fig. \ref{spatial_profile_KG_soft} (e) and (f).

On the other hand, hybrid stable large amplitude breathers can also exist in the gaps of the linear modes band, being denoted as phantom breather~\cite{phantom}; their properties depend on the lattice size.
The details of the above scenario are fairly similar in the hard potential case. Moreover, the dynamical evolution leads to a similar extended,
disordered redistribution of the lattice energy as discussed in the
DNLS case, and hence relevant illustrations are omitted here for brevity.

\section{A concrete realization proposal: a tailored electric circuit to observe spatially linear and nonlinear EM states}

\begin{figure}[h]
	\centering
	\includegraphics[width=\columnwidth]{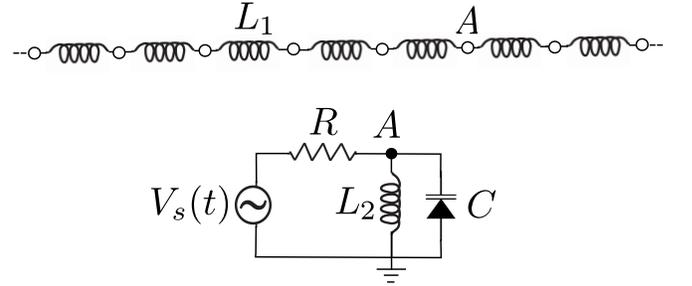}
	\caption{Schematic circuit diagrams of the electrical lattice,
where the circles  represent circuit cells. Each cell is connected to a periodic voltage
source $V_s(t)$ via a resistor R and grounded. Each point A of an
elemental circuit is connected via inductors $L_1$ to the corresponding
points A of neighboring cells. Voltages are monitored at point A.}
	\label{lineal_electric}
\end{figure}

The realm of electrical lattices has a time-honored history as
a highly controllable experimental testbed for the consideration
of localized modes~\cite{lars1,lars2}, as well as numerous variants
thereof, such as traveling~\cite{lars3}, higher-dimensional
realizations~\cite{lars4}, connections with the Kuramoto model~\cite{lars5},
and considerations of impurities~\cite{lars6}, among many others.
Here, we propose an electric line composed by resonant elements ($LC$ oscillators) coupled by inductors as shown in Fig. \ref{lineal_electric}.
This system is rather analogous to the ones used for the above
mentioned experimental setups and can be tailored to the needed properties.
To induce the excitation we need an external voltage source and a load resistor. The load resistor and the external voltage source will, with the right phase, allow us to induce the localized state. Once the localized state takes place we will turn off the driving and the load resistance. Thus, we could  observe, for a suitable time interval (given the dissipative effects), this state.

For example, in line with with parameter values that have been used
in some of the above experimental works, we will consider values
of capacitance $C = 770$ pF, inductances $L_1=680 \mu$H, $L_2=330 \mu$H, resistance $R=10$k$\Omega$ and voltage $V_s(t)=V_d \cos(\omega t+\varphi_n).$ A chain size that is quite suitable for experimental observations,
in line with the above works, 
is $N=31$ with free boundary conditions.
The (dimensionless) linear modes are the solutions of the equation
\begin{equation}
    \ddot{v}_n=\frac{L_2}{L_1}(v_{n+1}+v_{n-1}-2v_n)-v_n,
\end{equation}
hence this creates the possibility to induce a spatially localized linear mode by using a set of impurities $\{\epsilon_n\}$, as an effective potential
enabling the existence of an EM.
To build the relevant circuit  we have to replace inductors $L_2$ with a set of inductors $L_2^{(n)}$ whose values depend on the site, so as to create
the effective local potential. Thus, in this case:
\begin{equation}
    L_2^{(n)}=\frac{L_2}{1+\epsilon_n}.
\end{equation}

\begin{figure}[h]
	\centering
	\includegraphics[width=\columnwidth]{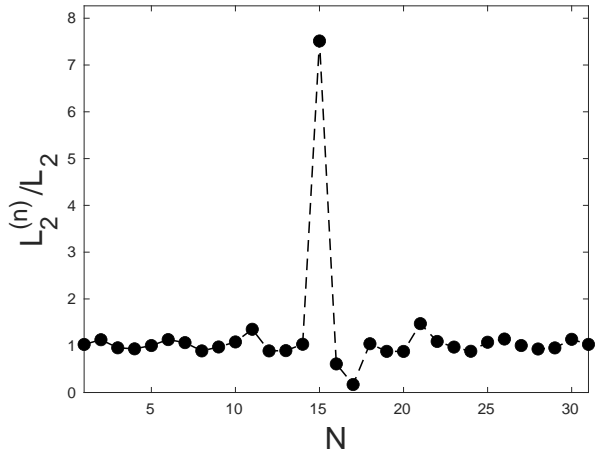}
\caption{Linear case: Spatial profile of the impurities $L_2^{(n)}$. $N=31$, $n_0=16$, and $\omega'=1.6453 \omega_0$, where $\omega_0=1/\sqrt{C_0 L_2}$}
\label{L}
\end{figure}

\begin{figure}[h]
	\centering
	\includegraphics[width=\columnwidth]{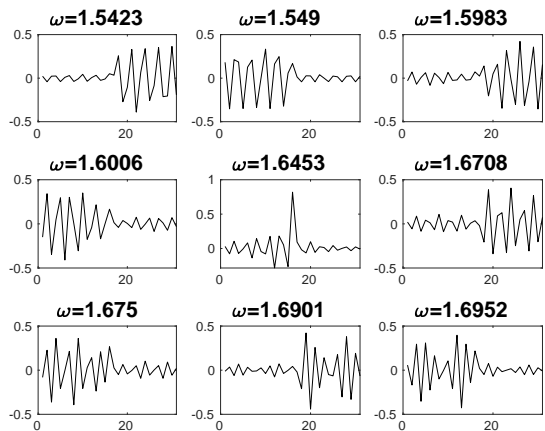}
	\caption{ Linear case: Spatial profiles of states in the band closest in frequency to the localized mode (central panel). $N=31$, $n_0=16$, and $\omega'=1.6453\ \omega_0$, where $\omega_0=1/\sqrt{C_0L_2}$.}
	\label{linear_electric_profile}
\end{figure}

In figure \ref{L} we show different values of inductors to obtain a linear localized state as shown in Fig. \ref{linear_electric_profile}.
By replacing the capacitor by a varactor it is possible to explore nonlinear localized modes in the phonon band. In principle, we believe that this concrete proposal
renders the phenomenology presented herein accessible to the current
state-of-the-art in electrical lattice experiments. This is true not
only at the linear level described above, but also at the nonlinear
one for suitably larger values of the voltage.

\section{Conclusions and Future Challenges}

In the present work, we have explored the possibility of realizing
embedded modes in the continuous spectrum of nonlinear dynamical lattices.
Leveraging earlier ideas from the realm of linear problems, 
we have systematically
constructed such modes in the linear (low-intensity) limit of the nonlinear
problem. Then, the use of nonlinearity (focusing or defocusing, soft or
hard) has allowed us to continue the relevant linear state in the presence
of nonlinearity at arbitrary strengths detuning away from the linear limit.
We have found that such states can arise as nonlinear extensions both
at the solitary wave setting of the DNLS model and  in the discrete breather
realm of nonlinear Klein-Gordon dynamical lattices. Moreover, in
the vicinity of the linear limit the states may be stable, while they
destabilize, typically towards spreading their mass throughout the
lattice, further away from the limit. Not only have we explored the
existence, stability and dynamics of such EM states, but we have also
proposed a concrete experimental implementation thereof in the form
of a 31-node electrical lattice of inductors and capacitors (and resistors,
and external drive, as is typical in such experiments).

It will of course be interesting to attempt to pursue such lattices
in future experimental work and to attempt to explore the damped-driven
variants of the relevant models (as in such experiments it is not
straightforward to completely eliminate dissipation towards the Hamiltonian
realm as in this study).

Another interesting direction is to
examine generalizations of such modes in higher dimensions and the
exploration of how higher-dimensionality may affect the stability
conclusions reached herein. As yet another interesting point, considering such EMs close to the band edge of the continuous
spectrum and examining how their stability may be affected by this
effective ``distance'' in frequency from the band edge would be another
interesting property to explore in the context of such nonlinear embedded
modes. These topics are currently under investigation and associated
findings will be reported in future publications.

\section*{Acknowledgements}
{M. I. M. acknowledges support by FONDECYT Grant 1200120. J.C-M. thanks the Regional Government of Andalusia under the project P18-RT-3480 and MICINN, AEI and EU (FEDER program) under the project PID2019-110430GB-C21. F.P. acknowledges financial support from the
Ministerio de Ciencia, Innovaci\'{o}n y Universidades (MICIU, Spain) through
Project No. PID2019-108508GB-I00 cofinancied by FEDER funds.
This material is based upon work
supported by the US National Science Foundation under Grant  
DMS-1809074 (PGK).}


\end{document}